\documentclass[12pt,a4paper]{article}
\usepackage[utf8]{inputenc}
\usepackage[margin=2cm]{geometry}
\usepackage{amsmath,amssymb}
\usepackage{graphicx}
\usepackage{bm}
\usepackage{array}

\newcommand{\dd}[2]{\frac{d #1}{d #2}}

\title{Modelling oral adrenal cortisol support}
\author{David J.\ Smith\(^{12}\)\thanks{d.j.smith@bham.ac.uk}, Alessandro Prete\(^{23}\), Angela E.\ Taylor\(^{23}\),\\ Niki Karavitaki\(^{23}\) and Wiebke Arlt\(^{23}\)}
\date{\small \(^1\)School of Mathematics \& \(^2\)Institute for Metabolism and Systems Research,\\ University of Birmingham, Edgbaston, Birmingham, B15 2TT, UK,\\[0.5em] \(^3\)Centre for Endocrinology, Diabetes and Metabolism, Birmingham Health Partners,\\ Birmingham, B15 2GW, UK.}

\begin{document}

\maketitle

\begin{abstract}
    A simplified mathematical model of oral hydrocortisone delivery in adrenal insufficiency is described; the model is based on three components (gastric hydrocortisone, free serum cortisol and bound serum cortisol) and is formulated in terms of linear kinetics, taking into account the dynamics of glucocorticoid--protein binding. Motivated by the need to optimise cortisol replacement in the situations of COVID-19 infection, the model is fitted to recently-published data on 50~mg dosing and earlier data on 10~mg dosing. The fitted model is used to predict typical responses to standard dosing regimes, which involve a larger dose in the morning and 1 or 2 smaller doses later in the day, and the same regimes with doses doubled. In all cases there is a circadian-like response, with early morning nadir. The model is also used to consider an alternative dosing strategy based on four equal and equally-spaced doses of 10, 20 or 30 mg per 24 h, resulting in a more even response resembling a response to sustained inflammatory stress.
\end{abstract}

\section*{Background}

The Prevention of Adrenal Crisis in Stress (PACS) study \cite{prete2020} aimed to identify the parenteral hydrocortisone dose and administration mode most suitable for glucocorticoid stress dose cover in patients with adrenal insufficiency exposed to major stress, such as trauma, surgery or sepsis. This included experimental data on serum cortisol levels measured by mass spectrometry after administering 200mg hydrocortisone in four different administration modes (50~mg qds orally or via bolus intramuscular or intravenous injections, and continuous intravenous infusion of 200~mg/24~h). That paper included a model of intravenous hydrocortisone administration and clearance, from which predictions could be made, identifying an initial bolus of 50mg or 100~mg hydrocortisone followed by continuous intravenous infusion of 200~mg hydrocortisone per 24 hours as the most appropriate intervention.
 
Patients with adrenal insufficiency are required to increase their usual oral hydrocortisone dose when experiencing intermittent illness with fever, which usually involves doubling of their regular glucocorticoid replacement dose, with twice the regular dose taken at the same timepoints as usual. Further deterioration then requires switching to parenteral hydrocortisone replacement for major stress dose cover as described above. During the time of COVID-19, however, patients with adrenal insufficiency might require higher oral stress doses already initially, as the viral illness caused by SAYS-cov2 often comes with significant fever, sweating and malaise early on. In addition, the high and frequently continuous fever requires a more sustained delivery of hydrocortisone that adjusts in dose and timing to the permanent inflammatory stress. Thus, we used the experimental data on oral hydrocortisone administration \cite{prete2020} together with building on a previously developed approach \cite{bunte2018} to adapt the model for oral hydrocortisone administration, enabling us to select the most suitable dose to recommend.

\section*{Model of oral administration with binding kinetics}

We now describe an idealised model of oral hydrocortisone treatment. Comparing with the 1-component linear kinetics model of Prete et al.\ \cite{prete2020} for intravenous delivery, it is necessary to take into account the presence of a gastric compartment from which hydrocortisone must be absorbed, then transported to the blood. A model consisting of two compartments only (gastric and serum cortisol) was initially attempted, however it failed to replicate the dose response characteristics as an oral dose is increased from 10~mg to 50~mg. We therefore expanded the model to take into account a limited bound component, modelling the effect of binding protein, which slows excretion at lower doses.

In detail the components of the model are: gastric hydrocortisone dose (\(S(t)\)~mg, which is increased by \(Q_j\) each time a dose is taken), free serum cortisol \(F_f(t)\)~nmol/L and bound cortisol \(F_b(t)\)~nmol/L. The concentration of binding protein is accounted for via \(B(t)\). Reactions will be modelled as linear in all cases, however the finite quantity of binding protein available will result in a nonlinear response.

The reactions in the model are:
\begin{itemize}
    \item Uptake from stomach to blood, at rate \(k_{abs}\) and with dilution factor \(\alpha\).
    \item Excretion of free cortisol at rate \(k_{ex}\). Bound cortisol is assumed not to undergo significant excretion \cite{sarkar2013}.
    \item Binding of free cortisol to binding protein at rate \(k_{b}\) and 
    \item release of bound cortisol and protein at rate \(k_{r}\).
\end{itemize}

The system takes the form,
\begin{align}
    \dd{S}{t} & = -k_{abs}S + q(t), \\
    \dd{F_f}{t} & =  \alpha k_{abs}S - k_{ex} F - k_b F_f B + k_r F_b, \\
    \dd{F_b}{t} & = k_b F_f B - k_r F_b, \label{eq:fb}\\
    \dd{B}{t} & = - k_b F_f B + k_r F_b, \label{eq:b}
\end{align}
where \(q(t)\) is a function modelling the oral dosing. The initial conditions for an adrenal insufficient patient will be approximated as \(S(0)=0\), \(F_f(0)=0\), \(F_b=0\) and \(B(0)=B_0\) where \(B_0\) is the physiological level of binding protein (a value quoted in the literature is 650~nmol/L \cite{wood1994}, although this will vary between individuals).

The case of oral administration of a \(N\) doses \(\bm{Q}=(Q_1,Q_2,\ldots, Q_N)\) at times \(\bm{t}=(t_1, t_2, \ldots, t_N)\) respectively can be represented by a sum of Dirac delta functions,
\begin{equation}
    q(t) = \sum_{n=1}^N Q_d\delta(t-t_n).
\end{equation}

The differential equaiton for \(S(t)\) can be integrated, yielding,
\begin{equation}
    S(t) = \exp(-k_{abs}t) \int_{t_0}^t q(t') \exp(k_{abs}t') dt',\label{eq:Ssol}
\end{equation}
where starting time \(t_0\) precedes the first dose. Denoting the heaviside function by \(H\), equation~\ref{eq:Ssol} can be evaluated as,
\begin{equation}
    S(t) = \sum_{n=1}^N Q_n H(t-t_n) \exp(k_{abs}(t_n-t)).
\end{equation}

Adding equations~\eqref{eq:fb} and \eqref{eq:b} shows that \(F_b(t)+B(t)\) is constant and hence equal to its initial value of \(B_0\). Therefore the variable \(F_b(t)=B_0-B(t)\) may be eliminated from the model, leading to the two-variable system,
\begin{align}
    \dd{F_f}{t} & = \alpha\sum_{n=1}^N Q_n H(t-t_n) \exp(k_{abs}(t_n-t)) - k_{ex}F_f - k_b F_f B + k_r (B_0-B(t)), \label{eq:Ff2}\\
    \dd{B}{t} & = -k_b F_f B + k_r (B_0-B)\label{eq:B2}.
\end{align}

The fraction of free to total cortisol \(F_f/(F_f+F_b)\) is typically 5\% \cite{mattos2013}. Working on the assumption that the binding/unbinding processes occur faster than absorption and excretion leads to the quasi-steady approximation
\begin{equation}
    0\approx -k_b F_f B + k_r (B_0-B),
\end{equation}
which can be justified formally via dimensional analysis. Then
\begin{equation}
    F_f \approx \left(\frac{k_r}{k_b B_0}\right) (B_0-B) \frac{B_0}{B}.
\end{equation}
The first dimensionless parameter grouping will be denoted \(\phi:=k_r/(k_b B_0)\); this grouping quantifies the relative size of free to bound cortisol and will be found through fitting to time-course data. The reason for working with this parameter grouping is that the full model is insensitive to the value of \(k_b\), meaning that it is challenging to fit directly. We then have the approximation
\begin{equation}
    F_f\approx \phi (B_0-B) \frac{B_0}{B},\label{eq:qsapprox}
\end{equation}
which enables the system to be reduced to a single variable.

Noting that the total cortisol is given by \(F_f+F_b=F_f+(B_0-B)\), we may write
\begin{equation}
    \dd{(F_f+F_b)}{t}=\dd{(F_f-B)}{t},
\end{equation}
then subtracting equation~\eqref{eq:B2} from \eqref{eq:Ff2} yields,
\begin{equation}
    \dd{(F_f-B)}{t}=\alpha k_{abs}S-k_{ex}F_f.
\end{equation}
Applying equation~\eqref{eq:qsapprox} then leads to
\begin{equation}
    \dd{}{t}\left(\phi(B_0-B)\frac{B_0}{B}-B\right) = \alpha k_{abs} S - k_{ex}\phi(B_0-B)\frac{B_0}{B},
\end{equation}
hence
\begin{equation}
    \dd{B}{t} = \left(1+\phi\frac{B_0^2}{B^2}\right)^{-1}\left( - \alpha k_{abs} S + \phi k_{ex} B_0\left(\frac{B_0}{B}-1\right)\right).
\end{equation}

This model can then be solved numerically for \(B(t)\), from which the total cortisol concentration \(F(t)=F_f(t)+F_b(t)\) is then given by equation~\eqref{eq:qsapprox}. There are three remaining parameters to estimate by fitting to time series data for \(F(t)\): the dilution factor \(\alpha\), absorption rate \(k_{abs}\) and excretion rate \(k_{ex}\).

\section*{Parameter estimation}

Two datasets are used for fitting the model: the PACS \cite{prete2020} data on oral administration of 50~mg every 6~h in primary adrenal insufficiency patients, and published data from ref.\ \cite{debono2009} on oral administration of a single dose of 10~mg in HPA-suppressed healthy individuals. Data from the latter study were extracted digitally from the electronic journal article. The parameters \(\alpha\), \(k_{abs}\), \(k_{ex}\), \(\phi\) and \(B_0\) are fitted to the two datasets simultaneously. 

\begin{figure}
    \centering
    \includegraphics{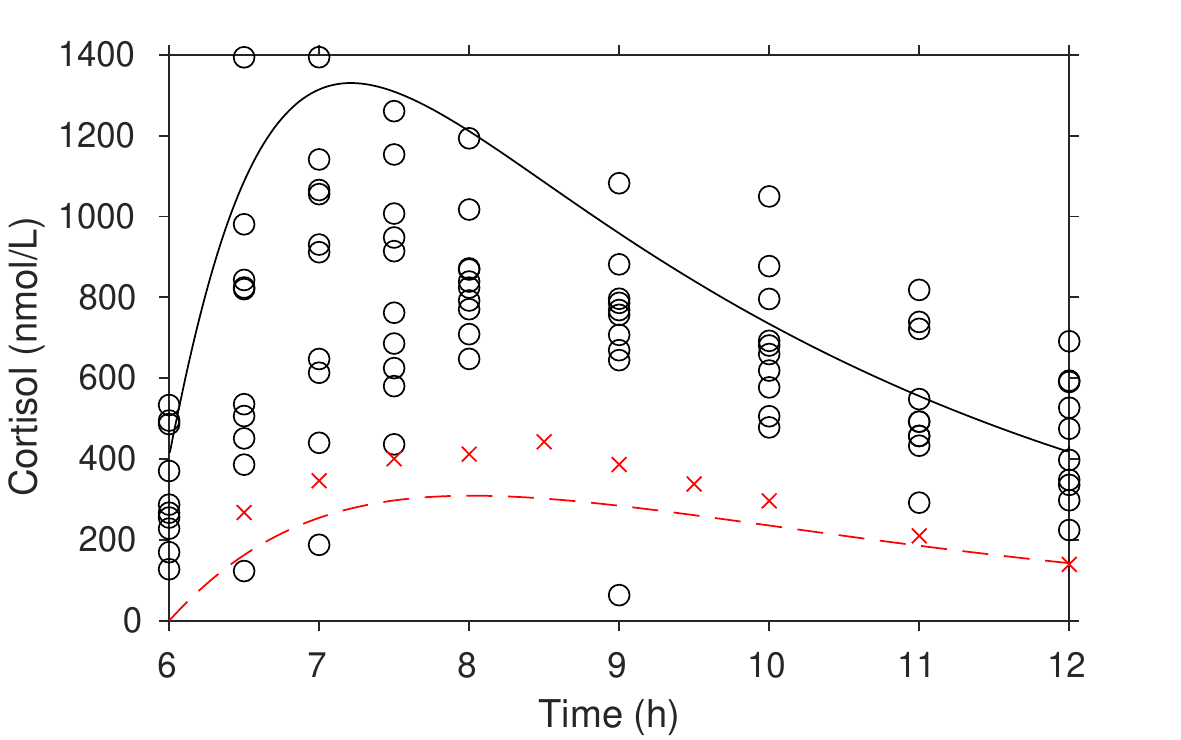}
    \caption{Oral dose data for cortisol (black: 50 mg every 6 hours \cite{prete2020}; red: 10 mg \cite{debono2009}) during the \(6\)--\(12\)~h interval and associated 3-compartment model simultaneous fit to the data. Time axes are aligned so that the dose commences at the time designated \(t=6\). Parameter estimates are  \(\alpha=69\)~(nmol/L)/mg \(k_{ex}=1.7\)~h\(^{-1}\), \(k_{abs}=0.59\)~h\(^{-1}\), \(\phi=0.23\) and \(B_0=1026\)~nmol/L.}
    \label{fig:fitting}
\end{figure}

A high degree of fidelity is unlikely in this situation due to attempting to fit to two different patient groups, dosing formulations and assays \emph{simultaneously} with a relatively simple model; figure~\ref{fig:fitting} shows reasonably good agreement, although appearing to under-predict the response at 10~mg and over-predict at 50~mg. Nevertheless, the model does appear to be capable of giving approximate information over a 5-fold change in dose.

\begin{table}
    \centering
    \begin{tabular}{cc>{\raggedright}p{2.75cm}c}
        Short name & Summary & Detail & Parameters \\\hline
        15--10 & standard regime I  & 15 mg at 7:00, 10 mg at 13:00 & \(\bm{t}=(7,13)\), \(\bm{Q}=(15,10)\)\\
        10--5--5 & standard regime II & 10 mg at 7:00, 5 mg at 13:00, 5 mg at 17:00 & \(\bm{t}=(7,13,17)\), \(\bm{Q}=(10,5,5)\) \\
        30--20     & double standard regime I  & 30 mg at 7:00, 20 mg at 13:00 & \(\bm{t}=(7,13)\), \(\bm{Q}=(30,20)\) \\
        20--10--10 & double standard regime II & 20 mg at 7:00, 10 mg at 13:00, 10 mg at 17:00 & \(\bm{t}=(7,13,17)\), \(\bm{Q}=(20,10,10)\) \\
        10--10--10--10 & 10 mg every 6 h &  10 mg at 0:00, 6:00, 12:00, 18:00 & \(\bm{t}=(0,6,12,18)\), \(\bm{Q}=(10,10,10,10)\) \\
        20--20--20--20 & 20 mg every 6 h &  20 mg at 0:00, 6:00, 12:00, 18:00 & \(\bm{t}=(0,6,12,18)\), \(\bm{Q}=(20,20,20,20)\) \\
        30--30--30--30 & 30 mg every 6 h &  30 mg at 0:00, 6:00, 12:00, 18:00 & \(\bm{t}=(0,6,12,18)\), \(\bm{Q}=(30,30,30,30)\)
    \end{tabular}
    \caption{Dosing regimes used to produce the model results in figure~\ref{fig:predictions}.}
    \label{tab:regimes}
\end{table}

\begin{figure}
    \centering
    \begin{tabular}{ll}
    \raisebox{7cm}{(a)} & 
    \includegraphics{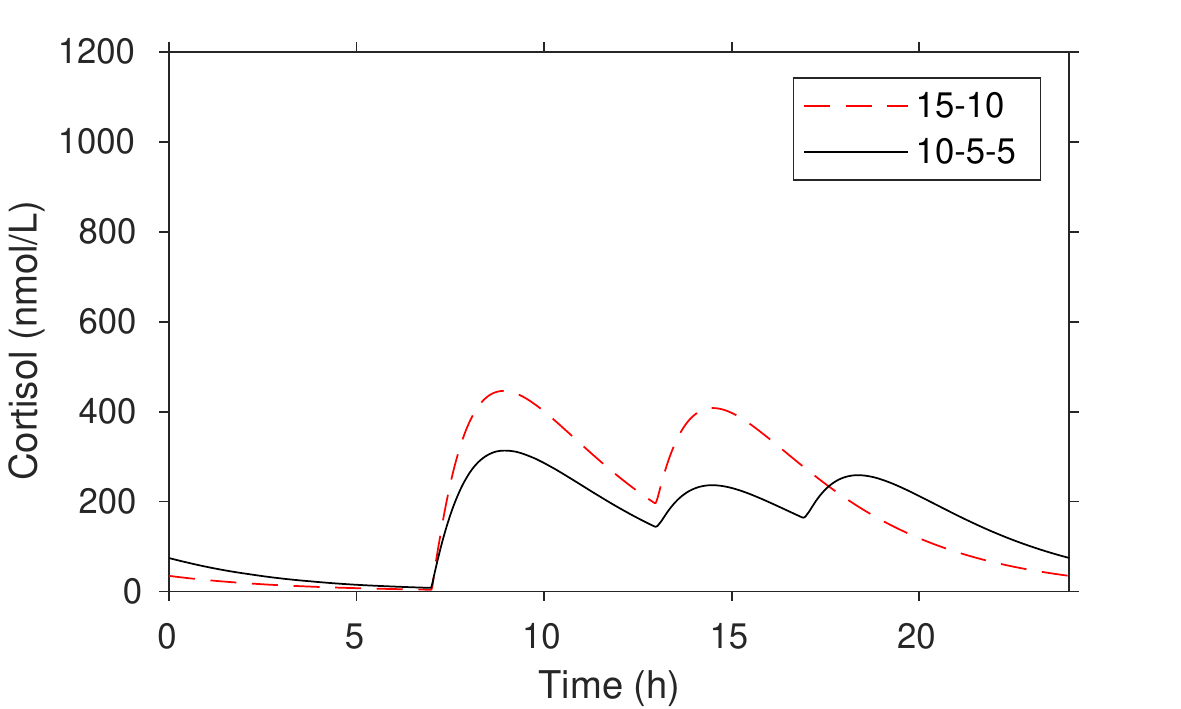} \\
    \raisebox{7cm}{(b)} & 
    \includegraphics{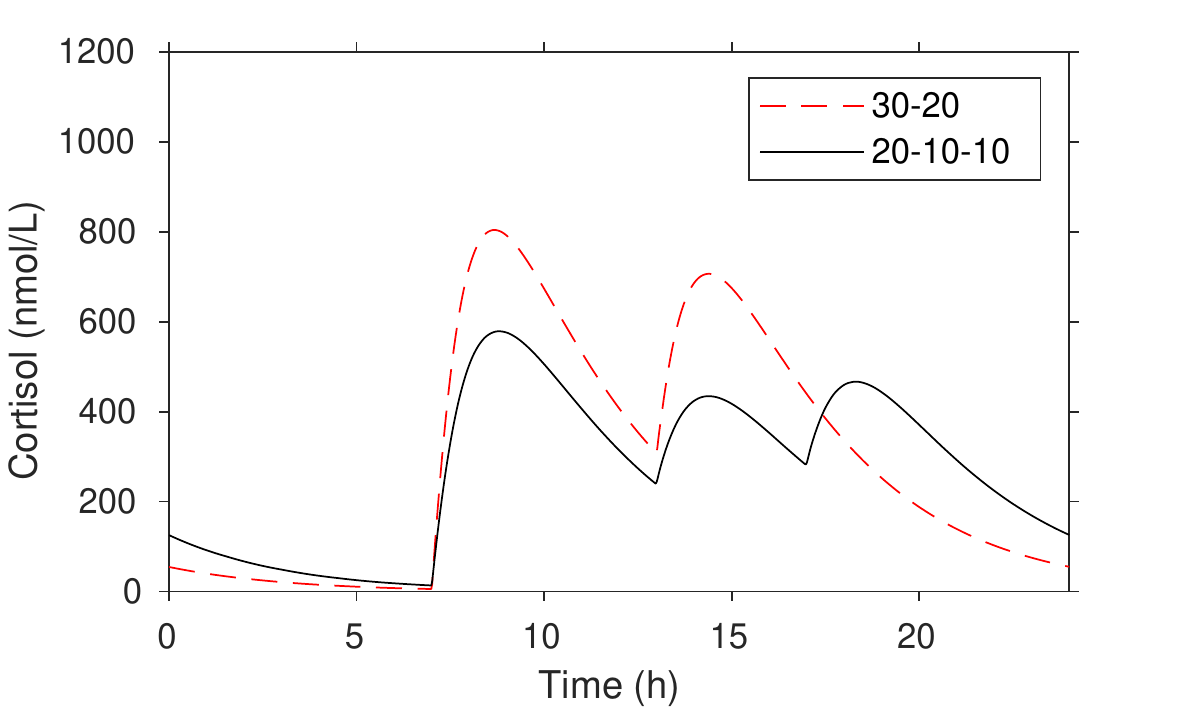} \\
    \raisebox{7cm}{(c)} & 
    \includegraphics{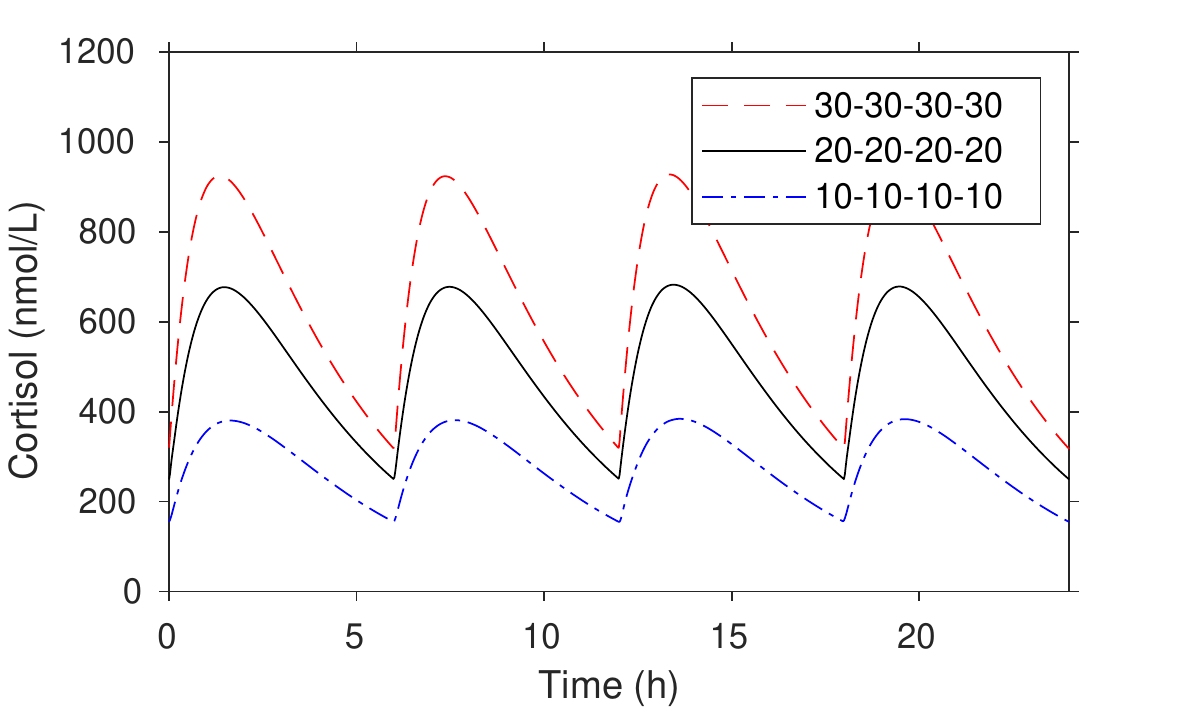}
    \end{tabular}
    \caption{Serum cortisol model predictions based on the parameter estimates in figure~\ref{fig:fitting}. (a) Standard regimes, (b) doubled standard regimes, (c) even-spaced doses. Details are given in table~\ref{tab:regimes}. In all cases dosing was commenced in the 24 hour period prior to the period plotted to enable a regular pattern of response to establish.}
    \label{fig:predictions}
\end{figure}

The model can then be used to predict typical responses to a range of dosing regimes, as detailed in table~\ref{tab:regimes}. Figure~\ref{fig:predictions}(a,b) show the predicted time courses for the standard (regime I: 15 mg at 7:00 followed by 10 mg at 13:00; regime II: 10 mg at 7:00 followed by 5 mg at 13:00 and 5 mg at 17:00), and doubled regimes, which replicate a circadian-like pattern with nadir just before the first dose at 7:00, and maintained concentrations through most of the day from around 7:30 to midnight. 

Motivated by the aim to replicate the adrenal response to continuous physiological stress, figure~\ref{fig:predictions}(c) shows the predicted time courses with 6-hourly equal doses of each of 10, 20 and 30 mg. In all cases the low cortisol nadir is avoided.

\section*{Model limitations and discussion}
The model described here is somewhat idealised and does not attempt to replicate in detail multiple physiological compartments, heterogeneity between patients nor characteristics such as age, sex and ethnicity: more detailed pharmacokinetic-pharmacodynamic models (such as ref.\ \cite{melin2018}) would give more detail and information around variability in responses. The fitted model appears to underpredict peak responses at 10~mg and over-predicts at 50~mg, in each case by around 25\%. Oral dosing is also subject to first-pass liver metabolism, which metabolise steroids for downstream excretion, and indeed the PACS study showed higher excretion of hydrocortisone metabolites in urine than equivalent parenteral doses (\cite{prete2020}, suppl.\ fig.\ 2). 
It is also possible that physiological response to COVID-19, in particular body temperature, may affect rates in the system such as binding globulin affinity \cite{chan2013}. Results should be interpreted with these limitations in mind.

Nevertheless, the model provides an approximation of the shape of the 24 hour time-courses which would result from the dosing strategies considered. Dosing every 6~h clearly indicates that an early morning nadir in cortisol can be avoided, which may be valuable for patients suffering from the stress of viral infection.

\section*{Funding}
Data were generated via the \emph{Prevention of adrenal crisis in surgery} study \cite{prete2020}, supported by Medical Research Council UK (program grant G0900567), the Oxfordshire Health Services Research Committee, and the National Institute for Health Research  (NIHR)  Birmingham  Biomedical  Research  Centre  at  the  University  Hospitals  Birmingham NHS Foundation Trust and the University of Birmingham (grant reference number BRC-1215-2009). AP is a  Diabetes  UK  Sir  George  Alberti  Research  Training  Fellow  (grant  reference number  18/0005782).

The  views  expressed  are  those  of  the  authors  and  not  necessarily  those  of  the  NIHR  or  the Department  of  Health  and  Social  Care  UK. 

\section*{Acknowledgements}
The authors thank Profs.\ Richard Ross and Simon Pearce for valuable discussions.

\end{document}